\documentstyle[psfig,aps,prl,multicol]{revtex}

\newcommand{\vx}{\vec{v}}
\newcommand{\va}{\vec{a}}
\newcommand{\be}{\begin{equation}}
\newcommand{\ee}{\end{equation}}

\begin{document}
\title{Identifying and modelling delay feedback systems}
\author{Rainer Hegger, Martin J.\ B\"unner~\thanks{currently at:
Istituto Nazionale di Ottica, Largo E.\ Fermi 6, 50125 Firenze,
Italy}, Holger Kantz}
\address{Max--Planck--Institut f\"ur Physik komplexer Systeme\\
N\"othnitzer Str.\ 38, 01187 Dresden, Germany}
\author{Antonino Giaquinta}
\address{Dip.\ di Ingegneria dei Sistemi e Informatica\\
via S.\ Marta 3, University of Florence, 50125 Firenze, Italy}
\date{\today}
\maketitle

\begin{abstract}
Systems with delayed feedback can possess chaotic attractors with
extremely high dimension, even if only a few physical degrees of
freedom are involved. We propose a state space reconstruction
from time series data of a scalar observable, coming along with a novel
method to identify and model such systems, if a single variable is fed
back. Making use of special properties of the feedback structure, we
can understand the structure of the system by
constructing equivalent equations of motion in spaces with dimensions
which can be much smaller than the dimension of the chaotic
attractor. We verify our method using both numerical and experimental
data. 
\end{abstract}

\begin{multicols}{2}

Since it is known that simple nonlinear dynamical systems can produce
extremely complicated, aperiodic motion, the search for
low-dimensional deterministic structures in observed time series has
become a common approach to the understanding of complex time
dependent phenomena. The methods, summarised under the term 
non-linear time series analysis, have meanwhile acquired 
a high standard\cite{Kantz97,Abarbanel93b}. It is possible, in spite
of problems with noise or non-stationarity,
to extract estimates of the fractal dimensions,
the entropy and the Lyapunov exponents from observed data. Often, the
ultimate goal is the construction of model-equations from
the data.

As a first step of this kind of data analysis, a state space has to be
reconstructed from the data, since usually only a single observable is
measured (scalar data)\cite{Packard}. The mathematical basis for this
reconstruction is given by the delay embedding theorems of
Takens~\cite{Takens80} and Sauer et
al.~\cite{Sauer91a}. The theorem, in the formulation of the latter
authors, states that, loosely speaking, forming $m$-dimensional
delay vectors $\vx_i=(x_i,x_{i-1},\ldots,x_{i-m+1})$ of the sequence
of measurements $\{x_j\}$ yields a one to one description of the
attractor of the system, if $m>2D_f$, where $D_f$ is the fractal
dimension of the attractor. If $D_f$ is larger than about 5, this
method usually fails in practice, since the number $N$ of observed
measurements covers the high dimensional attractor
insufficiently. Additionally, the time delay embedding method
introduces distortions for data with large Kolmogorov-Sinai entropy,
such that the typical minimal inter-point-distance which can be reached
by reasonably large data sets (say, $N=10^6$) is still too large for a
detection of the deterministic structure~\cite{Kantz97a}.  

A special class of dynamical systems is that of the {\it delayed
feedback} systems~\cite{Hale}. Despite a small number of physical
variables, their phase space is infinite dimensional, namely
the space of all differentiable functions on the time interval
$[0,\tau_0]$, where $\tau_0$ is the delay time of the feedback. They
can thus possess chaotic attractors of arbitrarily high dimension. The
direct reconstruction of the attractor by the time delay embedding
method is therefore usually
impossible. Recently\cite{Buenner96,potsdam}, a method
was proposed to identify such systems from data when they involve only
a single variable, i.e. when they can be described by a scalar
delay-differential equation
\be
\dot{x}(t)=f(x(t),x(t-\tau_0))\;.
\label{eq.tdde}
\ee
The structure of the evolution
equation~(\ref{eq.tdde}) is obviously independent of the dimension of the
resulting attractor. Indeed, in the space spanned by the
vectors $\left(\dot{x}(t),x(t),x(t-\tau)\right)$,
Eq.~(\ref{eq.tdde}) defines a constraint if $\tau=\tau_0$ and the data
collapse onto a two dimensional manifold, defined by
$\dot{x}(t)-f(x(t),x(t-\tau_0))=0$. This property was used
in~\cite{Buenner96,potsdam} to identify these systems and to recover
the delay time $\tau_0$ and the function $f$. Similar results were
achieved for $d$-dimensional systems if all $d$ components are
measured~\cite{Buenner97}.

In this paper we introduce a nontrivial generalisation
for delay systems with more than one variable in the realistic situation
that only a single observable of the system was measured. This method
allows one to reconstruct state vectors
with which we can determine the time delay $\tau_0$ and an equation of
motion. We will thus be able to perform predictions and to model the system.
The idea of our method is to combine a Takens--like embedding technique
with the structure of time delayed feedback systems, and to use the
ability of making forecasts as a
criterion for a successful reconstruction of a state space. Thus the
state space reconstruction
implies the derivation  of evolution equations in this space.

We start with the one-dimensional case~(\ref{eq.tdde}). In the
following,  $\tau_0$ will be  
the time delay of the feedback measured in multiples of the sampling
interval $\Delta t$ and for simplicity will be assumed to be integer,
whereas $\tau$ will be a trial value used in the reconstruction. A
successful reconstruction will lead to $\tau=\tau_0$.
We make the ansatz
\be
\hat{\dot{x}}_i=g_i(x_i,x_{i-\tau})\;,
\label{eq.ansatz}
\ee
where $\hat{\dot{x}}_i$ is an estimate for the time derivative
computed from differences of the time series data $\{x_j\}$. 
For $g_i$ we choose a local linear model
$g_i(\vx_i)=b_i+\va_i\vx_i$, where the parameters $b_i$ and $\va_i$ depend
on the spatial coordinates $\vx_i=(x_i,x_{i-\tau})$. This model represents
the first terms of a Taylor series expansion of the unknown global
nonlinear model around $\vx_i$. 
The parameters can be obtained for
each time index $j$ by the local least squares fit
\be
\sigma^2_j(\tau)=\frac{1}{N_{U_j}}\sum_{\vx_i\in U_j}
\left(\hat{\dot{x}}_i-g_j(x_i,x_{i-\tau})\right)^2\;,
\ee
where $U_j$ is a sufficiently small neighbourhood of
$\vx_j=(x_j,x_{j-\tau})$ 
and $N_{U_j}$ its cardinal number. The correct unknown value
$\tau_0$ of Eq.(1) is recovered if $\sigma(\tau)=\langle
\sigma_j(\tau)\rangle_j$ is at its absolute minimum, since only for
this value of $\tau$ all triples $(\hat{\dot{x}}_i,x_i,x_{i-\tau})$
fulfil simultaneously a constraint, if the motion is chaotic. 

Figure~\ref{fig.mg1-ospe} shows the relative forecast errors for
the Mackey--Glass~\cite{Mackey77} system
\be
\dot{x}=\frac{ax(t-\tau_0)}{1+x^{10}(t-\tau_0)}-bx(t)\;,
\label{eq.mg-def}
\ee
with $a=0.1$, $b=0.2$ and $\tau_0=100$. The figure demonstrates that
the correct delay time is detected easily. 
Once we know $\tau_0$, we can determine $g_j$ to iterate the
system by a simple Euler integration $x_{j+1}=x_j+\Delta t g_j(\vx_j)$,
provided the numerical trajectory remains on the attractor formed by
the observed data. It is plausible that
one could as well replace the local linear models $g_i$ by a single
global nonlinear function $g(\vx)=\sum a_j f_j(\vx)$, 
e.g.\ a bivariate polynomial or radial basis functions, by minimising
$\sum_{i=2}^N (x_{i+1}-\sum_j a_j f_j(x_i,x_{i-\tau}))^2$ with respect to the
coefficients $a_j$, which is standard in nonlinear function
fitting. We used the radial basis functions successfully for the
Mackey-Glass data discussed above.

In general one has to expect a delayed feedback system to involve
several variables. For the relevant situation that only one of these
variables enters the feedback loop the equation of motion reads
\be\label{eq.moredim}
\dot{\vec{y}}(t)=\vec{f}\left(\vec{y}(t),y^{\ell}(t-\tau_0)\right)\;,
\label{eq.d-dim.def}
\ee 
where $\vec{y}\in{\bf R}^d$ and $y^{\ell}(t-\tau_0)$ is the single component which
is fed back into the system. In many laboratory experiments with time
delayed feedback exactly this is the case. 
If we again possess only a scalar time series, we have to recover
the unobserved variables of the system in 
addition to determining the delay time $\tau_0$. 
We can do this by exploiting
the ideas of Casdagli for input--output systems~\cite{Casdagli92}.

Consider a non-autonomous deterministic system with an arbitrary input
$\epsilon_t$ (which could even be noise) of the form
\be
\vec{y}_{t+1}=\vec{f}(\vec{y}_t,\epsilon_t)\;,
\ee
where $\vec{y}\in{\bf R}^d$. If we possess a scalar time series of
one component of $\vec y$, a standard 
time delay reconstruction cannot work, since each
additional measurement 
contains an additional uncertainty, namely $\epsilon_t$. If
one simultaneously measures the input, one can form delay vectors
\be
\vec{v}_{t}=\left(y_t,y_{t-1},\ldots,y_{t-m+1},\epsilon_t,
\ldots,\epsilon_{t-m+1}\right)\;.
\ee
Casdagli argues that in this space the deterministic dynamics can be
reconstructed, if $m=2d$, generically~\cite{Casdagli92}.

In our case the delayed feedback variable can be interpreted as
non--autonomous input. Following Casdagli's reasoning, we have to include
these inputs in the delay vector,
\be 
\vec{v}_i(\tau)=\left(x_i,x_{i-1},\ldots,x_{i-m_1+1},x_{i-\tau},\ldots,
x_{i-\tau-m_2+1}\right)\;.  
\ee 
Here it is assumed that $x$ is identical to or a unique function of
only the delayed variable from Eq.~(\ref{eq.moredim}), and
$m_1$ is the dimension we need to fully determine the actual state of
the system, while $m_2$ is the number of inputs required. In general
we expect $m_2=m_1$. Due to strong correlations between successive
measurements $m_2<m_1$ could be sufficient. Following Casdagli's results
we expect that $m_1 = m_2=2d$ is sufficient for a
unique representation of $(\vec y(t),y^{\ell}(t-\tau))$, $t=i\Delta t$, of
Eq.~(\ref{eq.moredim}). Note that the fractal
dimension of the attractor can be much larger than $4d$ and that, in
particular, the embedding we propose is independent of the attractor
dimension. 

In analogy to the 1-dimensional case we make an ansatz in
this embedding space to recover the delay time and the dynamics of the
$d$-dimensional system:
\be
\hat{\dot{x}}_i=g_i(\vec{v}_i(\tau))\;,
\label{eq.m-dim-ansatz}
\ee
where $g$ here is again a local linear model, namely
$g_i(\vec{v}_i(\tau))=b_i+\vec{a}_i\vec{v}_i(\tau)$, and a successful
reconstruction will lead to $\tau=\tau_0$.

For a demonstration on numerically generated data we use a 
generalisation of the Mackey--Glass equation introduced in \cite{Buenner97}:
\begin{eqnarray}
\dot{x}(t)&=&\frac{ax(t-\tau_0)}{1+x^{10}(t-\tau_0)}+y(t)\label{eq.gen_mkg}\\
\dot{y}(t)&=&-\omega^2x(t)-\rho y(t)\nonumber\;,
\end{eqnarray}
where we choose the parameters to be $a=3$, $\rho=1.5$, $\omega^2=1$
and $\tau_0=10$. The 
Kaplan--Yorke dimension of this system is $D_{KY}\approx13.5$. The left
panel of Fig.~\ref{fig.mg-cast} shows its attractor  in a
two--dimensional representation. Figure~\ref{fig.mg2-ospe} shows the
one step prediction errors in an embedding space of the variable $x$
as a function of $\tau$. Already for an insufficient reconstruction 
$m_1=m_2=1$ (upper curve) we see indications of the correct
$\tau_0$. If we increase $m_1$
and $m_2$ to 2, the forecast errors decrease dramatically. At the
correct value of $\tau$ it is reduced by a factor of 20. Further
increases of $m_1$ and $m_2$ do not yield better forecast errors. 

This result can be verified analytically. 
An expression for the dynamics in delay space
follows from rewriting Eq.~(\ref{eq.gen_mkg}) as second order in time
\begin{eqnarray}
\ddot{x}(t)=&-&\omega^2 x(t)-\rho\dot{x}(t)+\omega^2f(x(t-\tau_0))\nonumber\\
&+&\frac{df(x(t-\tau_0))}{dx(t-\tau_0)}\dot{x}(t-\tau_0)\;,
\label{eq.thisisnoteq10}
\end{eqnarray}
where $f$ is the first term of the rhs of Eq.~(\ref{eq.gen_mkg}), and
substituting derivatives by differences. Using a simple 
Euler--scheme for this substitution, two 'on--time'
measurements  and two delayed measurements are involved in the
forecast, i.e.~$m_1=m_2=2$.
The right panel of Fig.~\ref{fig.mg-cast} shows data obtained by
iterating the fitted dynamics $g$ in
Eq.~(\ref{eq.m-dim-ansatz}). The properties of the
attractor are reproduced well, although the dimension of the
delay space is less than a third of the estimated Kaplan--Yorke
dimension. 

Experimental data from systems whose model equations are not fully
known are a challenge. We apply our state space reconstruction
scheme to data from a CO$_2$ laser experiment with
delayed feedback performed at the Istituto Nazionale di Ottica in
Firence/Italy~\cite{Arecchi86}. The output
power of the laser is measured, which is the variable which is
electronically fed back into the system.  Experimental data
are shown in the left panel of Fig.~\ref{fig.politi} in a
representation which was proposed in~\cite{Giacomelli94}. The scalar
time series $\{x_i\}$ is represented in a space--time plane by
decomposing the time index $i$ into an artificial
discrete time $n$ and a space variable $s$
\be
i=s+n\tau_0\quad\mbox{with}\quad s\in[0,\tau_0[\;,
\label{thisisneithereq10}
\ee
This presentation allows one to observe that the 
system behaves periodically with
defects travelling through it. Applying our analysis in 
$m_1=m_2=2$ dimensions we obtain a delay of $\tau_0=800$,
in perfect agreement with the value given by the experimentalists,
and the forecast error drops by a factor of $\approx10$. Enlarging
$m_1$ and $m_2$ does not improve the results, whereas $m_1=m_2=1$
turns out to be insufficient. Measurement noise may
here distort the estimates of the derivatives.
The right panel of Fig.~\ref{fig.politi} shows the
presentation of the time series obtained from the integrated fitted
equation of motion. Comparing the structures visually we see 
a good agreement, differing details are even to be expected for a
perfect model because of the sensitive dependence on initial conditions. 
We thus conclude that the experimental
data from the CO$_2$-laser are governed by a two-dimensional
time-delay differential equation with a single delayed feedback
variable. Therefore, we detect a considerable
reduction of dimension in comparison to the six-dimensional model 
proposed in~\cite{Varone95}, which has been derived from first
principles. Our finding is in agreement with the conjecture 
of~\cite{Giacomelli96} concerning the reduction of
the dimension of a time-delay system close to a bifurcation.
Without making use of the feedback structure, in standard 
embedding spaces with dimensions up to $6$, we were unable to obtain
models which reproduce the defect dynamics, and the optimal prediction
error was much larger.

When only a single quantity is measured in a system with time
delayed feedback, the full state space is not accessible and also the
``true'' equations of motion. We proposed a novel method for the
reconstruction of a space which is equivalent to the physical one.
In this space equations of motion attain a very
natural form, mirroring the delay structure of the unknown
dynamical system.
We thus work in a space whose dimension is usually 
much smaller than the attractor's dimension, although larger than the
number of variables in the unknown space. 
We do not reconstruct the {\sl attractor} by Takens's
embedding, but only the unobserved variables and then exploit
the special structure of the evolution equations. This is of
particular interest when $\tau_0$ in the experiment is varied: The
attractor dimension will vary, too, but the structure of 
our space remains invariant,
and so do the equations of motion in this space.
A successful reconstruction involves to detect the 
correct time delay $\tau_0$, which is done here very efficiently 
by a minimisation of a kind of prediction error. A maximum of the
auto-correlation function usually appears close to, but  not exactly
at $\tau_0$~\cite{Lepri93}.

Although we only 
presented results for systems with a single delay time, our method can
of course be easily generalised to systems with several delay
times. Furthermore, once we have modelled the system we can estimate
the Lyapunov spectrum from this model.

The problem of noise in experimental data will be subject to future
works. However, the successful treatment of the laser data shows that
well controlled laboratory experiments provide data whose noise
contamination does not harm the applicability of the concept. On the
other hand it is clear that the forecast error cannot be smaller than
the noise level, so that too strong noise could completely destroy the
detectability of the correct $\tau_0$.

We thank the experimentalists at the INO for their
excellent data. Moreover, we are indebted to R.\ Genesio, 
P.\ Grassberger, E.\ Olbrich, A.\ Politi and
T.\ Schreiber for very stimulating discussions. A.\ G.\ acknowledges
support of the EC under the contract number ERBFMRXCT960010 and
support from the MPIPKS.

\end{multicols}

\begin{figure}[t]
\caption{\protect\small One step prediction error for the
Mackey--Glass system with $\tau_0=100$.}
\label{fig.mg1-ospe}
\end{figure}

\begin{figure}[ht]
\caption{\protect\small One step prediction error for the
two--dimensional generalized Mackey--Glass system with $\tau_0=10$.}
\label{fig.mg2-ospe}
\end{figure}

\begin{figure}[ht]
\caption{\protect\small The attractor of the two dimensional
Mackey--Glass system. The left panel shows the original data, the
right panel the data obtained by iterating the fitted dynamics using
$m_1=m_2=2$.}
\label{fig.mg-cast}
\end{figure}

\begin{figure}[ht]
\caption{\protect\small Output power of a $CO_2$ laser experiment. The left
panel shows the original data, the right panel shows the data obtained
from an iterated forecast in $m_1=m_2=2$ dimensions. The
representation of the data is described in the text.}
\label{fig.politi}
\end{figure}

\end{document}